\documentclass[12pt]{iopart}
\pdfoutput=1
\usepackage[usenames,dvipsnames]{color}
\usepackage{graphicx}
\usepackage[utf8]{inputenc}
\usepackage[T1]{fontenc}
\usepackage{iopams}

\begin{document}

\title[Absolute negative mobility induced  by white Poissonian noise]
%Equivalence of white shot-noise and constant bias in the anomalous transport regime]
%{Controlling transport of inertial Brownian particles by Poissonian noise}
{Absolute negative mobility induced  by  white Poissonian noise}
%for  inertial Brownian particles
%Equivalence of Poissonian white shot noise and constant bias in the anomalous transport regime}
\author{J. Spiechowicz, J. {\L}uczka }
\address{Institute of Physics, University of Silesia, 40-007  Katowice, Poland}
\author{P. H{\"a}nggi}
\address{Institute of Physics, University of Augsburg,
%Universit\"atsstr. 1,
86135 Augsburg, Germany}

\ead{Jerzy.Luczka@us.edu.pl}

\begin{abstract}
	We research the transport properties of  inertial Brownian particles which move in  a \emph{symmetric} periodic potential and are subjected to  both a \emph{symmetric}, unbiased time-periodic external force  and biased  Poissonian white shot noise (of non-zero average $F$) being composed of a random sequence of  $\delta$-shaped pulses with random amplitudes. Upon varying the parameters of white  shot-noise one conveniently can manipulate the transport direction and the overall nonlinear response behavior. Within tailored parameter regimes, we find that the response is {\it opposite}  to the applied  average bias $F$  of such white shot noise. This very transport characteristics thus mimics a nonlinear Absolute Negative  Mobility (ANM)  regime.  Moreover,  such white shot noise driven ANM is  robust with respect to  statistics of the shot noise spikes. Our findings can be checked and corroborated experimentally by use of a  setup that consists of a single resistively and capacitively shunted Josephson junction device.
\end{abstract}


\maketitle

\section{Introduction}
\label{sec1}
	
In accordance with the Le Chatelier-Braun principle, when an external deterministic force $F$ acts on a particle with all other forces set to zero on average, it is expected that the long-time, stationary average particle velocity
$\langle v\rangle$  becomes an increasing function of the load $F$, at least for small bias values $F$. For an electrical (electronic) device, the current-voltage characteristics  exhibits a similar property:  if the voltage $V$  increases the current $I$ increases as well.  The Ohmic characteristics
$I=(1/R)\, V$ presents an example. This behavior  is usually characterized as  "normal transport" behavior.
In contrast, anomalous transport features are (i)  a negative differential mobility or conductivity,
(meaning that the velocity or current decrease with increasing force or voltage)
or (ii) an  absolute negative mobility (ANM) or conductivity,  i.e.
 the velocity or current exhibits an opposite sign to the applied force which is starting out at zero  force or voltage;
i.e. the system  response is  opposite to the applied force.

Such absolute negative mobility has been experimentally detected in a variety of systems, both in classical and quantum ones.  Typical situations where ANM has been detected are  p-modulation-doped GaAs quantum wells \cite{hop},  sequential resonant tunnelling semiconductor superlattices that are driven by intense tera-Hertz electric fields \cite{keay}, relaxing Xe-plasma ionized by a hard x-ray pulse \cite{warman},   sliding charge-density-waves  at sufficiently low temperatures \cite{zant},
in microfluidic systems for colloidal beads in an aqueous buffer solution
\cite{ros}, in a three-terminal configuration in a two dimensional
electron gas \cite{kaya},  or  transport of vortices in superconductors with inhomogeneous pinning under a driving force \cite{xu}
and  in Josephson junctions \cite{nagel}. Rather recently,  a  coherent absolute negative mobility regime has been observed and described for ac and dc driven ultra-cold atoms in an optical lattice \cite{sergiej}.

Yet further  examples  of ANM have been described theoretically for the
nonlinear response in ac-dc-driven tunnelling transport
\cite{HarGri1997},  the dynamics of cooperative Brownian motors
\cite{BroBen2000},  Brownian transport containing a complex
topology \cite{EicRei2002a,EicRei2002b} or  some
multi-state models with state-dependent noise \cite{CleBro2002}.
The effect of an absolute negative mobility can occur as well in driven systems such as
nonlinear inertial Brownian dynamics
\cite{prl_jj,prb_jj,acta_jj,reim},  overdamped nonlinear Brownian
motion in presence of time-delayed feedback \cite{hennig},   transport of
asymmetric particles  in a periodically segmented 2D-channel \cite{PeterH} and for a system of two coupled resistively shunted Josephson junctions \cite{jan}.

The key ingredient for the occurrence of ANM in all those listed cases is that the system (i) is driven far
away from thermal equilibrium into a time-dependent nonequilibrium state and the resulting dynamics is so that (ii)  it  is exhibiting a vanishing, unbiased average non-equilibrium response. In presence of a finite bias $F$,  the ANM response in a symmetric priodic potential is so that an average, anti-symmetric transport velocity $<v(F)>$, obeying $<v(F)> = - <v(-F)>$ occurs around the zero bias force $F=0$. This situation must be contrasted with the nonequilibrium transport generated by the rachet mechanism \cite{RMP2009}: There, a nonvanishing transport velocity occurs even for vanishing bias $F=0$, and thus no anti-symmetric mobility behavior occurs around the zero bias regime. The reader is thus advised to carefully distinguish between anomalous mobility regimes that are characterized  either as a negative-differential mobility regime, an absolute negative mobility regime or as a nonlinear negative-valued mobility away from the zero bias regime \cite{prb_jj}.

With this work, we substitute the deterministic static force $F$ by a random force $\eta(t)$ of non-zero average.
To allow for the comparison with the case of a deterministic
load $F\ne 0$, we require that the mean value of the random force $\eta(t)$	equals the  value $F$.  As a model for such a  random force we use  a random sequence of exponentially distributed $\delta$-shaped pulses with random amplitudes. This constitutes a generalized white Poissonian shot noise process; e.g. see in \cite{HPZfB78} for its detailed statistical properties.  We demonstrate  that this  shot noise can induce ANM and in the regime of  ANM,  it behaves on average statistically similar to  a deterministic constant bias $F$.  Moreover,  the ANM-phenomenon  is robust with respect to   the distribution of the random  amplitudes of the  $\delta$-pulses.
	
	The layout of the present work is as follows. In section \ref{sec2} we detail the model of a driven inertial Brownian particle. In section  \ref{sec3} we detail more closely the  stochastic force acting on the particle and elucidate the resulting  transport properties. The findings are contrasted   with an equivalent setup consisting of a deterministic bias. Section \ref{sec4} provides our summary and some conclusions.

%%%%%%%%%%%%%%%%%%%%%%%%%%%%%%%%%%%%%%%%%%%%%%%%%%%%%%%%%%%%%%%%%%%%%%%%%%%%%

\section{Model}
\label{sec2}
In what follows we  consider an ensemble of classical, statistically independent Brownian particles undergoing transport in an effectively one-dimensional geometry. For such a one-dimensional system, the {\it minimal}  model of the classical Brownian particle exhibiting the ANM  is formulated in terms of the equation of motion  for
a particle of mass $M$ moving in  (i) a symmetric spatially periodic potential $V(x)= V(x+L)$ of period $L$, (ii) being driven by an unbiased time-periodic  force $A \cos(\Omega t)$ with angular frequency $\Omega$ and amplitude strength $A$,  and (iii) exposed to a static force $F$. All three components are essential in order to take the system away from a thermal equilibrium state into a time-dependent driven  nonequilibrium state so that the limiting Le Chatelier-Braun  principle no longer applies. The corresponding Langevin equation reads  \cite{prl_jj}
\begin{equation}
	\label{mini}
		M\ddot{x} + \Gamma\dot{x} = - V'(x) + A\cos(\Omega t) + F + \sqrt{2\Gamma k_BT}\, \xi(t).
	\end{equation}
Here, the dot and the prime denotes the differentiation with respect to time $t$ and the Brownian particle's coordinate $x$, respectively. The parameter  $\Gamma$ is the friction coefficient, $k_B$ is the Boltzmann constant and $T$ is  the temperature.
Thermal equilibrium fluctuations are modelled by $\delta$-correlated Gaussian white noise $\xi(t)$ of zero mean and unit intensity, i.e.
	\begin{equation} \label{gauss}
		\langle \xi(t) \rangle = 0, \quad \langle \xi(t)\xi(s) \rangle = \delta(t-s).
	\end{equation}
We are interested in  the asymptotic long time regime, where the
 averaged velocity  assumes the periodicity of the driving \cite{jung1990}, i.e.,
\begin{eqnarray}
 \label{aver}
\langle v \rangle = \lim_{t\to\infty} \frac{\Omega}{2\pi}
\int_{t}^{t+2\pi/\Omega} \prec v(s) \succ \; ds,
\end{eqnarray}
where $\prec v(s) \succ$ indicates the average over noise realizations
(ensemble-average).  In the deterministic case ($T=0$), an additional average over initial conditions must be performed.

From the symmetry of the Langevin  equation it follows that the transformation $F \to  - F$ implies $\langle v \rangle \to -\langle v \rangle$. In other words, the average velocity $\langle v \rangle(F)$  as a function of the load $F$ fulfills the relation:
$\langle v \rangle(-F) =  -\langle v \rangle (F)$. In particular, it follows that the transport vanishes identically $\langle v \rangle \equiv 0 $ for $F=0$. This is in clear contrast to the case of a ratchet mechanism which exhibits finite transport at vanishing static bias \cite{RMP2009}. Generally, the averaged velocity $\langle v \rangle$ is a nonlinear function of the bias $F$. However, for small values of $F$ one can expect that
a linear response regime is present and $\langle v \rangle$ assumes a linear function of the small bias; i.e.,
\begin{eqnarray}
 \label{mobil}
\langle v  \rangle = \mu F.
\end{eqnarray}
In the normal transport regime, the mobility coefficient $\mu$ is positive, $\mu > 0$; in distinct contrast,
 $\mu < 0$ for ANM.  In Ref. \cite{prl_jj}, it has been shown that the above system exhibits ANM and there are two fundamentally different mechanisms
for ANM: (a) induced by thermal fluctuations and (b)  generated by deterministic dynamics.

We  now substitute the deterministic static force $F$ by a stochastic force  $\eta(t)$.
Such a case is important because it could help to explain and clarify the understanding of unusual transport properties not only in physical but also in biological systems such as, e.g., the bi-directionality of
the net cargo transport inside living cells, where there are no a systematic deterministic load but rather random collisions in the form of kicks and impulses. Thus, in place of equation (\ref{mini}) we shall consider the setup
	\begin{equation}
	\label{modeleq}
		M\ddot{x} + \Gamma\dot{x} = - V'(x) + A\cos(\Omega t) + \eta(t) + \sqrt{2\Gamma k_BT}\, \xi(t).
	\end{equation}
	 The potential $V(x)$ is assumed to be in the simplest \emph{symmetric} form
	\begin{equation}	
		V(x) = \Delta V\sin(2\pi x/L).
	\end{equation}	
In order to make the comparison with the case of the deterministic
load $F\ne 0$, we set the mean value of the random force $\eta(t)$	equal to  $F$; i.e. $\langle \eta(t) \rangle = F$.
%It means that the random force  $\eta(t)$ exerts a net force on the particle.
As a model for such a  stochastic biased forcing  we propose a random  sequence of delta-shaped pulses with
random amplitudes defined in terms of
generalized  white Poissonian  shot noise  \cite{HPZfB78,hanggi}:
	\begin{equation}
	\label{pwsn}
		\eta(t) = \sum_{i=1}^{n(t)}z_i \delta(t - t_i),
	\end{equation}
	where $t_i$ are the arrival times of a Poissonian counting process $n(t)$ with parameter $\lambda$. Put differently, the probability for occurrence  of $k$ impulses in the time interval $[0, t]$ is governed by the Poisson distribution; i.e.,
	\begin{equation}
		\mbox{Pr}\lbrace n(t) = k \rbrace = \frac{(\lambda t)^k}{k!}e^{-\lambda t}.
	\end{equation}	
	Likewise, the interval $s$ between successive Poisson arrival times $s = t_i - t_{i-1}$ is exponentially distributed with the probability density $\lambda \exp(-\lambda s)$. The parameter $\lambda$ determines the mean number of the $\delta$-pulses per unit time or equivalently the mean spiking rate of the $\delta$-pulses. The amplitudes $\lbrace z_i \rbrace$ of the $\delta$-pulses denote independent random variables. These amplitudes are statistically distributed  according to a common probability density $\rho(z)$. The process $\eta(t)$ presents  white  noise of  finite mean  and a covariance  given by
	\begin{equation}
	\label{pwsnmom}
		\langle \eta(t) \rangle = \lambda \langle z_i \rangle, \quad \langle \eta(t)\eta(s) \rangle  - \langle \eta(t) \rangle \langle \eta(s) \rangle
		= 2D_P\delta(t-s),
	\end{equation}
	where $\langle z_i \rangle$ is an average over the amplitude distribution $\rho(z)$.
	Poissonian  white noise is statistically symmetric if the density  $\rho(z)$ is  symmetric, i.e.  when $\rho(z) = \rho(-z)$. Consequently,
	$\langle \eta(t) \rangle = 0$. However, we shall not consider unbiased driving but instead consider biased white
    Poissonian noise for which $\langle \eta(t) \rangle \ne  0$.  The white  Poissonian  noise intensity $D_P$  reads
	\begin{equation}
	\label{pwsnintensity}
		D_P = \frac{\lambda \langle z_i^2 \rangle}{2}.
	\end{equation}
	We further assume that thermal equilibrium noise $\xi(t)$ is uncorrelated with  non-equilibrium noise $\eta(t)$, so that
	$\langle \xi(t)\eta(s) \rangle = \langle \xi(t)\rangle \langle \eta(s) \rangle  = 0$.
	
	Next we use a dimensionless form of equation (\ref{modeleq}). This can be achieved in several ways: Here  we propose the use of the period $L$ as a length scale and  for time  the scale  $\tau = L\sqrt{M/\Delta V}$  \cite{luczka}, respectively. Consequently, equation \eref{modeleq} can be rewritten in  the dimensionless form as:
	\begin{equation}
	\label{dimlessmodeleq}
		\ddot{\hat{x}} + \gamma\dot{\hat{x}} = - \hat{V}'(\hat{x}) + a\cos(\omega \hat{t}) + \hat{\eta}(\hat{t}) + \sqrt{2\gamma D_G}\, \hat{\xi}(\hat{t}),
	\end{equation}
	where $\hat{x} = x/L$ and $\hat{t} = t/\tau$. Other re-scaled dimensionless parameters are the friction coefficient $\gamma = \tau \Gamma/M$, the amplitude $a = LA/\Delta V$ and  the angular frequency $\omega = \tau \Omega$ of the time-periodic driving. The rescaled  potential $\hat{V}(\hat{x}) = V(L\hat{x})/\Delta V = \sin(2\pi \hat{x})$ possesses the unit period:  $\hat{V}(\hat{x}) = \hat{V}(\hat{x} +1)$. We introduced the dimensionless thermal noise intensity $D_G = k_BT/\Delta V$, so that the  Gaussian white noise of vanishing mean $\hat{\xi}(\hat{t})$ possesses the auto-correlation function $\langle \hat{\xi}(\hat{t})\hat{\xi}(\hat{s}) \rangle = \delta(\hat{t} - \hat{s})$. Similarly, the re-scaled Poissonian white shot noise is  $\delta$-correlated as well; i.e.,  $\langle \hat{\eta}(\hat{t})\hat{\eta}(\hat{s}) \rangle -
\langle \hat{\eta}(\hat{t})\rangle\langle\hat{\eta}(\hat{s}) \rangle	
	= 2\hat{D}_P\delta(\hat{t} - \hat{s})$, with $\hat{D}_p = \hat{\lambda} \langle \hat{z}_i^2 \rangle /2$, where $\hat{\lambda} = \tau \lambda$ and $\hat{z}_i = z_i/\sqrt{M\Delta V}$. Hereafter, we will use only  dimensionless variables and shall omit the notation "hat" in all quantities appearing in equation  \eref{dimlessmodeleq}.
	
	The deterministic dynamics corresponding to equation \eref{dimlessmodeleq}  exhibits an extremely rich and complex behaviour. Depending on the parameter values  periodic, quasiperiodic and chaotic motion can be observed
\cite{jungkissnerhanggi}. In some regimes, ergodicity is broken  and
the direction of the spontaneous transport depends on the choice for the initial conditions.
Different initial conditions of position and velocity may lead to radically different asymptotic behaviour; i.e. various attractors may coexist. The asymptotic regime can be classified as either being a locked or a running state.
%The former is realized when the motion is confined to a finite number of spatial periods.
The regime of the running state is the crucial ingredient for the occurrence of non-vanishing transport in the deterministic regime, allowing the system to explore all of space. At non-zero temperature, the system will be typically  ergodic with  thermal fluctuations enabling  diffusive transport with stochastic escape events connecting coexisting deterministic disjunct attractors  \cite{hanggitalknerborkovec}. In particular, transitions between neighboring locked states give rise to diffusive  directed transport.
	
	There exist a wealth of physical systems that can be described by equations of the form in   \eref{dimlessmodeleq}. An important  case that comes to mind is the semi-classical dynamics of a phase difference across a resistively and capacitively shunted Josephson junction which is driven by both a  time-periodic and a random force \cite{kautz}. For this setup the space coordinate of the Brownian particle $x$ and the driving force  correspond to the phase difference and the current applied to the Josephson junction, respectively. Other specific systems are rotating dipoles in external fields \cite{rotdip}, superionic conductors \cite{supioncond} or charge density waves \cite{chardenswav}, to name just a few.
	
%%%%%%%%%%%%%%%%%%%%%%%%%%%%%%%%%%%%%%%%%%%%%%%%%%%%%%%%%%%%%%%%%%%%%%%%%	

\section{Transport properties of a Brownian particle driven by white Gaussian and white Poissonian shot noise}
\label{sec3}

	 The Fokker-Planck-Kolmogorov-Feller master equation corresponding to the Langevin equation \eref{dimlessmodeleq}, cf. \cite{hanggi}, cannot be studied within closed analytical means. Consequently, we have to resort to comprehensive numerical simulations of the white Gaussian and white shot noise  driven Langevin dynamics. Details of the employed numerical scheme can be found in Refs. \cite{numerics1,numerics2}. We have chosen the time step to be $0.002 \cdot 2\pi/\omega$ and  used  initial conditions $\{x(0), \dot{x}(0)\}$ that are equally distributed over the interval $[0, 1]$ and $[-2,2]$, respectively (remember that the rescaled potential possesses the unit period). Noise averaging has been performed over $10^3 - 10^6$ different stochastic realizations and, additionally,  over one period of the external driving period $2\pi/\omega$. All numerical calculations have been done by use of a CUDA environment implemented on modern desktop GPU. This scheme allowed for a  speed-up  of a factor  of the order   $ \sim 100$ times as compared to a common present-day CPU method \cite{januszkostur}.
	
\begin{figure}[h]
		\centering
		\includegraphics[width=0.5\linewidth]{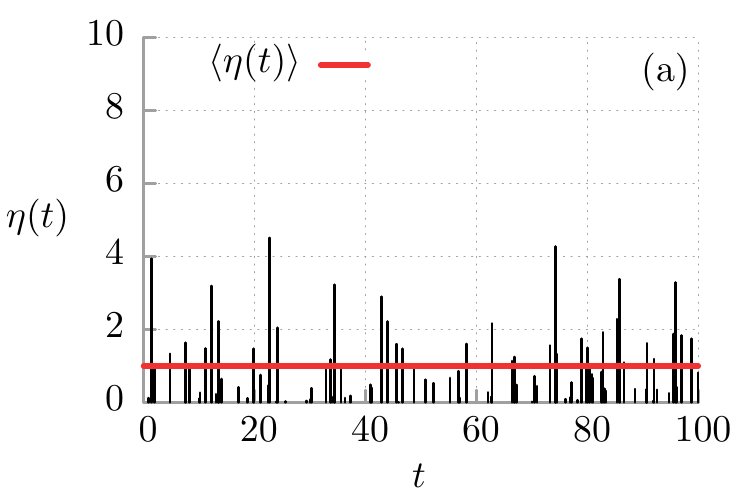} \\
		\includegraphics[width=0.49\linewidth]{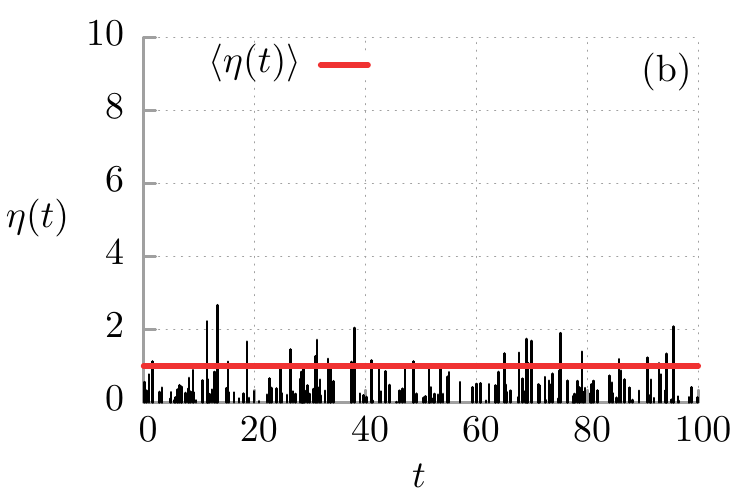}
		\includegraphics[width=0.49\linewidth]{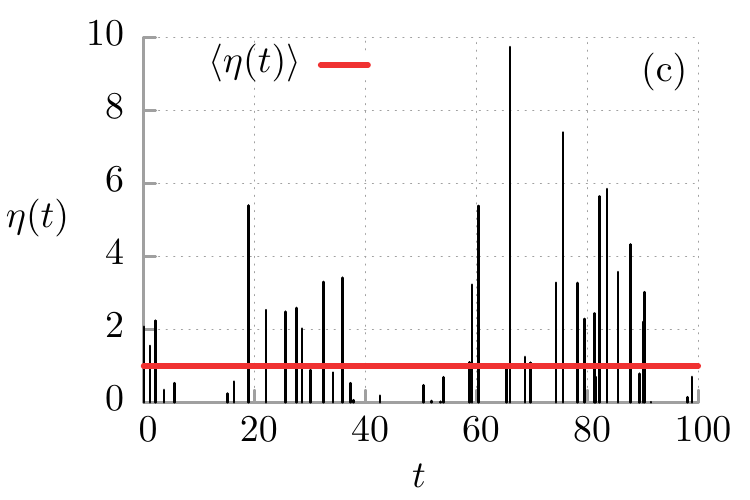}
		%\resizebox{0.49\linewidth}{!}{\input{tfig1a}} \\
		%\resizebox{0.49\linewidth}{!}{\input{tfig1b}}
		%\resizebox{0.49\linewidth}{!}{\input{tfig1c}}
		\caption{Three illustrative  realizations of Poissonian white shot noise $\eta(t)$. The amplitudes $\lbrace z_i \rbrace$ of the $\delta$-spikes are distributed according to the  exponential probability density (\ref{rho1}). In all three cases, the mean value is held fixed, i.e.,  $\langle \eta(t) \rangle = \zeta \lambda = \sqrt{\lambda D_P} =1$. The spike rate $\lambda$ and the noise intensity $D_P$ are varied as follows:  In (a): $\lambda = 1$, $D_P = 1$, in (b): $\lambda = 2$, $D_P = 0.5$, in (c): $\lambda = 0.5$, $D_P = 2$.}
		\label{fig1}
	\end{figure}

	To gain insight into the role of the white Poissonian shot noise we first examine the influence of the noise parameters $\lambda$ and $D_P$ on the characteristics of the stochastic realizations.  To be definite, we assume that the amplitudes $\lbrace z_i \rbrace$ of the $\delta$-kicks are exponentially distributed with the probability density
	\begin{equation}
	\label{rho1}
		\rho_1(z) = \zeta^{-1}\theta(z)e^{-z/\zeta},
	\end{equation}
	where $\theta(z)$ denotes  the Heaviside step function; i.e. the noise  amplitudes take on only positive values, $z_i > 0$. According to equation \eref{rho1}, the statistical moments of these amplitudes $\lbrace z_i \rbrace$  are given by
	\begin{equation}
	\label{rho1mom}
		\langle z_i^k \rangle = k!\zeta^k, \quad k = 1, 2, 3, ...
	\end{equation}
	From equations \eref{pwsnmom} and \eref{pwsnintensity} it follows that the mean value  $\langle \eta(t) \rangle$ and the  intensity $D_P$ of white  shot-noise read
	\begin{equation}
		\langle \eta(t) \rangle = \lambda \zeta, \quad D_P = \lambda \zeta^2.
	\end{equation}
	We next use $\lambda$ and $D_P$ as the quantifiers for transport; yielding   $\langle \eta(t) \rangle = \sqrt{\lambda D_P}$ and $\zeta = \sqrt{D_P/\lambda}$. Three typical realizations of white Poissonian noises are depicted in figure \ref{fig1}. In figure 1(a), the mean spiking rate of pulses $\lambda$ and the noise intensity $D_P$ are fixed at $1$. In order to ensure the condition $\langle \eta(t) \rangle = 1$ we may proceed in two different ways: A first one  is to increase the spiking rate
	$\lambda$  while reducing correspondingly the noise intensity $D_P$.
Then the  particle is {\it frequently}  kicked by	{\it weak} $\delta$-pulses.
	This case is depicted in figure \ref{fig1}(b). The second one is to  decrease  the spiking rate
	$\lambda$  and increase  the noise intensity $D_P$. Then the particle is {\it rarely} kicked by
	{\it large}  amplitudes of the $\delta$-spikes, cf. figure 1(c). -- It is worth mentioning that in the limit of vanishing  amplitudes $z_i$, when
$\zeta \to 0$ and a divergent spiking rate $\lambda \to \infty$, with
$ \lambda \zeta^2 = D_P$ held fixed,  the zero-mean process $\eta(t) - \langle \eta(t) \rangle$ approaches Gaussian white noise of  intensity $D_P$.

%%%%%%%%%%%%%%%%%%%%%%%%%%%%%%%%%%%%%%%%%%%%%%%%%%%%%%%%%%%%%%%%%%%%%%

\subsection{Ohmic-like transport regime}

	Let us  comment first on  the long time behaviour of the considered driven system in Eq.  \eref{dimlessmodeleq}. If the Poissonian  white shot noise is absent  in Eq.  \eref{dimlessmodeleq}, then the average velocity $\langle v \rangle$ is vanishing identically. This feature follows from  the presence of reflection symmetry  of the potential $V(x)$ and  the time reversal symmetry  of  harmonic driving. Put differently, a directed ratchet transport \cite{RMP2009}  is absent.

\begin{figure}[h]
		\centering
		%\resizebox{0.49\linewidth}{!}{\input{tfig2}}
		\includegraphics[width=0.49\linewidth]{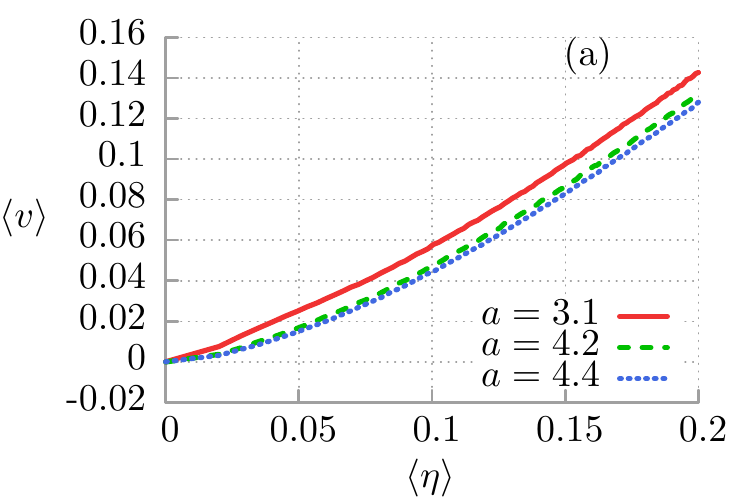}
		\includegraphics[width=0.49\linewidth]{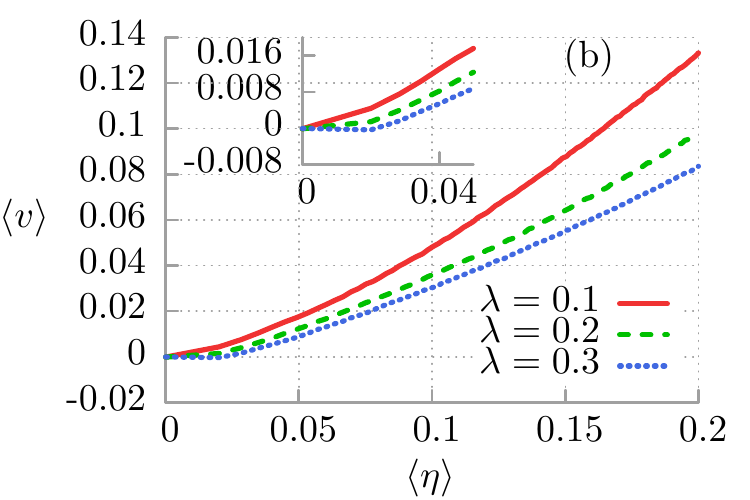}
		\caption{Ohmic-like  dependence of the asymptotic, time-averaged  asymptotic velocity $\langle v \rangle$ on the mean value of shot-noise $\langle \eta(t) \rangle$. Panel (a): the influence of amplitude strength $a$ of the time-periodic  driving is illustrated for  fixed spiking rate $\lambda = 0.1$ of $\delta$-spikes.
Panel (b): the role of the spiking rate is depicted for the ac-driving amplitude  $a=4.2$.  	
		The spike amplitudes $\lbrace z_i  \rbrace$ of the $\delta$-kicks are distributed according to the exponential probability density  $\rho_1(z)$ in equation  (\ref{rho1}).
		The remaining parameters are fixed as follows: the friction coefficient is $\gamma = 0.9$, the thermal fluctuation intensity $D_G = 0.001$ and  the angular driving frequency  is $\omega = 4.9$.}
		\label{fig2}
	\end{figure}

In the presence of white Poissonian  shot noise  driving, however, a statistical bias emerges \cite{epl2}, being due to the non-vanishing average.  This causes a   non-zero mean velocity, which typically assumes the  sign of the  average  of white shot-noise. We recall that there are only {\it positive} $\delta$-kicks and therefore we expect that average velocity is  positive as well.  An opposite behavior would be  counterintuitive. Because the dynamics as determined by equation  \eref{dimlessmodeleq} is strongly nonlinear and the stochastic phase space of the system is multidimensional, it should not come as surprise that the dependence of the average velocity on $\langle \eta(t) \rangle$ is  nonlinear as well and even may behave in a  non-monotonic manner of the system parameters. The normal expected behavior for the ensemble and time averaged   velocity is that of an increasing function for $\langle \eta(t) \rangle$. Such a normal  regime is depicted with Fig. \ref{fig2}. We used therein  the following parameter values: the friction coefficient  has been chosen as $\gamma = 0.9$, the thermal fluctuation intensity $D_G = 0.001$, the angular driving frequency is  $\omega = 4.9$. In panel (a), the influence of the  driving amplitude $a$ is shown. Panel (b) depicts the role of the spiking rate $\lambda$ of $\delta$-pulses.  The corresponding shot-noise characteristics corresponds  to rare   but large $\delta$-spikes. As a consequence the average velocity varies almost linearly on the mean value of shot-noise $\langle \eta(t) \rangle$, resulting in an Ohmic-like transport behavior.

%%%%%%%%%%%%%%%%%%%%%%%%%%%%%%%%%%%%%%%%%%%%%%%%%%%%%%%%%%%%%%%%%%%%%%%%%%%%

\subsection{Regime of Absolute Negative Mobility}

\begin{figure}[h]
		\centering
		\includegraphics[width=0.49\linewidth]{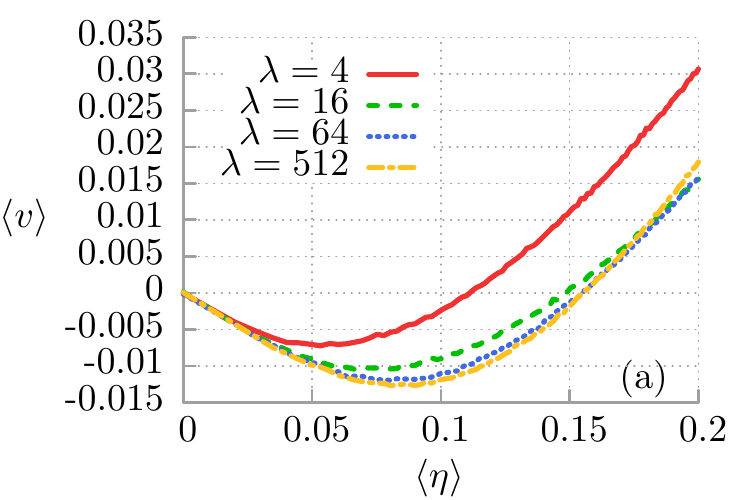}
		\includegraphics[width=0.49\linewidth]{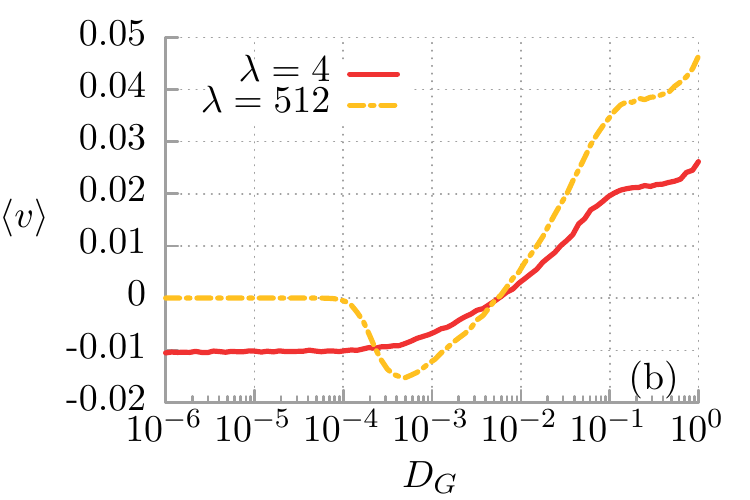}
		%\resizebox{0.49\linewidth}{!}{\input{tfig3a}}
		%\resizebox{0.49\linewidth}{!}{\input{tfig3b}}
		\caption{Absolute negative mobility (ANM). Panel (a):  The asymptotic time-averaged velocity $\langle v \rangle$ as a function of   the mean value of white shot-noise $\langle \eta(t) \rangle$ for various spiking frequencies $\lambda$ and fixed  thermal fluctuation intensity $D_G = 0.001$.  In panel (b) the role of thermal fluctuations is shown for two spiking frequencies $\lambda = 4$ (solid line, $D_P = 6 \cdot 10^{-4}$) and $\lambda = 512$ (dotted line, $D_P = 2 \cdot 10^{-5}$). Note that for the value $\lambda =512$, ANM is induced by thermal fluctuations. There occurs an optimal temperature $D_G$ at which  ANM  is most pronounced. In both panels: the amplitudes $\lbrace z_i  \rbrace$ of the  $\delta$-spikes  are exponentially distributed; the friction coefficient  $\gamma = 0.9$, the ac-driving amplitude   $a = 4.2$ with an angular driving frequency   $\omega = 4.9$. }
		\label{fig3}
	\end{figure}

	We have randomly searched the parameter space and  found that the normal,  Ohmic-like transport regime dominates in the parameter space.
Keeping in mind  that there are only positive $\delta$-kicks of white shot-noise acting  on the Brownian particle, we inquire  whether we can identify parameter regimes for which  the stationary mean velocity  of the particle assumes negative values, i.e. the particle moves on average opposite to the applied   $\delta$-spikes. In Fig. \ref{fig3}(a) we exemplify this situation. The characteristic feature is the emergence of extended regimes,
	$\langle \eta(t) \rangle >0$, where the average velocity $\langle v \rangle$, starting out from  zero, assumes a {\it negative} response; i.e. ANM occurs. Moreover, there exists an optimal  strength  for $\langle \eta(t) \rangle$ at which the average velocity assumes its  minimal value. We detect that if the spiking rate $\lambda$  increases,  the minimum of the resulting average transport velocity   is lowered.  Notably, we have found that there exists a limiting minimal value for  the transport velocity which is assumed for  $\lambda \to \infty$.  For $\lambda > 512$ (see the dotted line) the velocity characteristics becomes  numerically indistinguishable.
	%This limiting case coincides to the situation illustrated in figure \ref{fig1}b.
%If we replace shot-noise $\eta(t)$ by the constant force $F$ this behavior is termed the ANM \cite{prl_jj}: If $\eta(t) \equiv 0$ then $\langle v \rangle \equiv  0$. However,  for the non-negative random force,  $\eta(t) \ge 0$,  the averaged velocity is negative, $\langle v \rangle < 0$.
	
	The role of thermal fluctuations is depicted in panel (b). Two distinct mechanisms for the ANM-like effect can be observed:  For $\lambda = 4$,  the negative velocity is caused by deterministic chaotic dynamics because even at zero temperature $D_G=0$ the velocity is negative. In this regime,  temperature plays a destructive role for ANM: increase of  temperature monotonically diminishes the negative average velocity. For $\lambda = 512$, the negative velocity is   solely induced by thermal fluctuations. For low temperatures (small $D_G$) ANM does not occur. If thermal fluctuations grow ($D_G$ increases)
	the ANM effect emerges and intensifies up to optimal temperature where ANM
	is most pronounced. Subsequent increase of temperature reduces ANM and finally temperature destroys it completely.  For 'high' temperature, transport is normal.  The reader may find a detailed explanation of the origin of anomalous transport in such systems in Ref. \cite{prl_jj}.
	
%%%%%%%%%%%%%%%%%%%%%%%%%%%%%%%%%%%%%%%%%%%%%%%%%%%%%%%%%%%%%%%%

\subsection{Controlling transport}
	
\begin{figure}[h]
		\centering
		\includegraphics[width=0.49\linewidth]{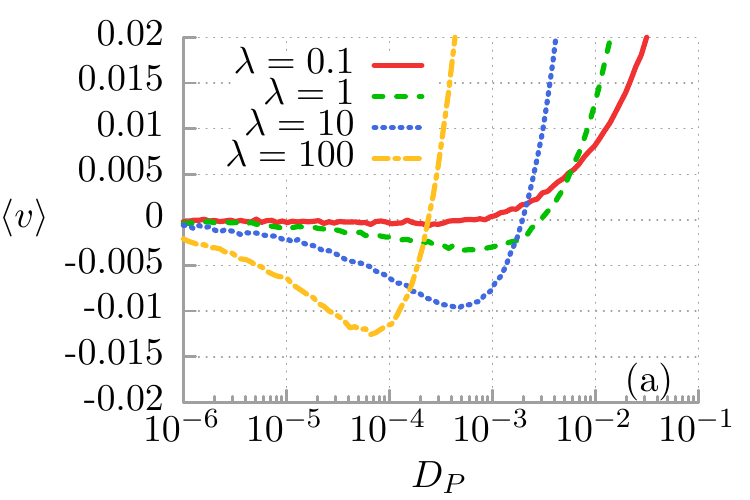}
		\includegraphics[width=0.49\linewidth]{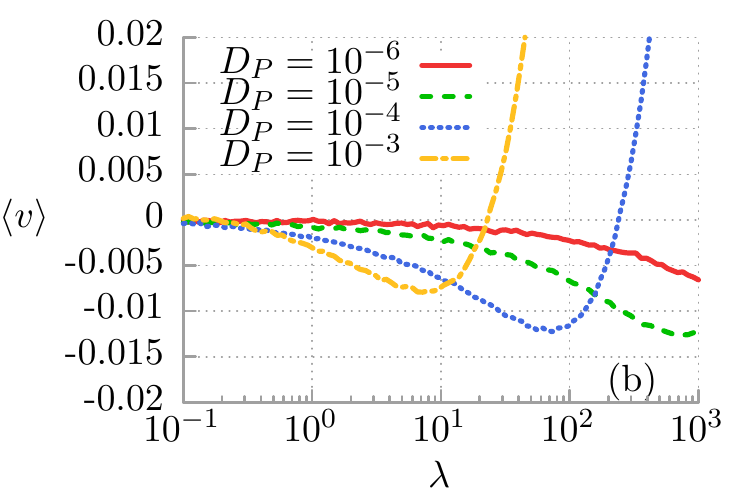}
		%\resizebox{0.49\linewidth}{!}{\input{tfig4a}}
		%\resizebox{0.49\linewidth}{!}{\input{tfig4b}}
		\caption{Panel (a):  The  time-averaged asymptotic velocity $\langle v \rangle$ as a function of  the white shot-noise intensity $D_P$ for selected values of spiking  frequencies $\lambda$. Panel (b): Average velocity as a function of the spiking rate $\lambda$ for several values of  white shot noise intensity $D_P$. The amplitudes $\lbrace z_i  \rbrace$ of the $\delta$-kicks are generated according to the distribution $\rho_1(z)$ and thermal fluctuation intensity $D_G = 0.001$. Other parameters are the same as those detailed in figure 3.}
		\label{fig4}
	\end{figure}

	The  transport properties  of the Brownian particle  can be  controlled by varying the parameters of white Poissonian noise; i.e., the values for  $\lambda$ or $D_P$. The dependence of the asymptotic average velocity $\langle v \rangle$ on these white shot noise   parameters is presented  in Figs. \ref{fig4}(a) and  \ref{fig4}(b).
	In Fig. 4(a)  we study the role of an increasing white shot noise intensity  $D_P$.  As an example, consider the
case with  $\lambda =10$ in Fig. 4(a).  For  weak shot noise intensity $D_P$ the velocity exhibits ANM
and its  minimal value decreases with  increasing  $D_P$. For strong white shot noise, however, when intensity $D_P$ is sufficiently large, the average transport  velocity turns around  towards a normal regime, undergoing a current reversal at some finite noise strength $D_P$.

The spiking rate $\lambda$ of the white shot noise as well serves as a control parameter for ANM.   The numerical findings are depicted with Fig.   4(b)  for a set of selected values of the white shot noise intensity $D_P$.
As before we  find that the ANM can be controlled upon varying the spiking rate  $\lambda$.  Again we detect a current reversal at finite spiking rate $\lambda$; this value of reversal  shifts to much larger  spiking frequencies  with decreasing white shot noise intensity $D_P$.
In summary,  one can  conveniently manipulate the direction of the particle transport  and  tune the ANM regime upon  varying the two shot noise parameters.

%%%%%%%%%%%%%%%%%%%%%%%%%%%%%%%%%%%%%%%%%%%%%%%%%%%%%%%%%%%%%%%%%%%%%

\subsection{Robustness of ANM on amplitude statistics}

\begin{figure}[h]
		\centering
		\includegraphics[width=0.5\linewidth]{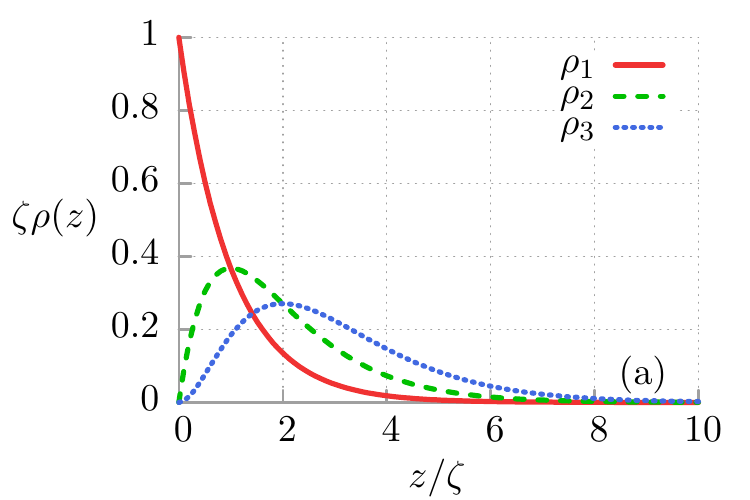} \\
		\includegraphics[width=0.49\linewidth]{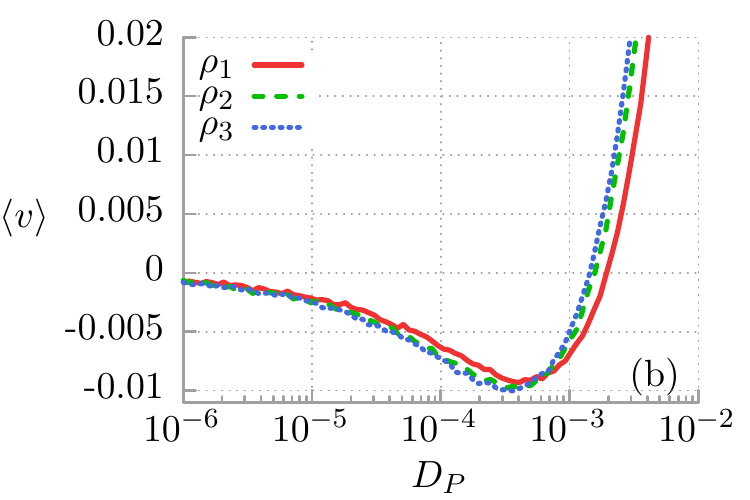}
		\includegraphics[width=0.49\linewidth]{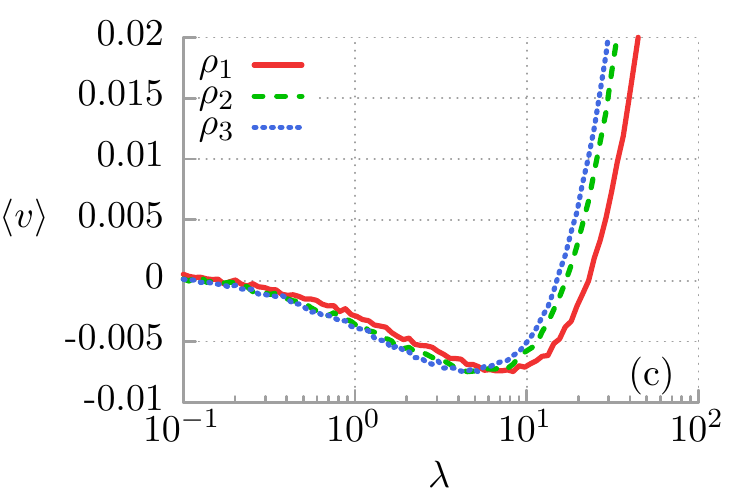}
		%\resizebox{0.49\linewidth}{!}{\input{tfig5a}} \\
		%\resizebox{0.49\linewidth}{!}{\input{tfig5b}}
		%\resizebox{0.49\linewidth}{!}{\input{tfig5c}}
		\caption{Panel (a): Plots of three probability densities of the amplitudes  $\{z_i\}$. Panel (b):  The dependence of the time-averaged, asymptotic velocity $\langle v \rangle$ on the intensity $D_P$  of shot-noise $\eta(t)$ for three statistical distributions of the amplitudes $\lbrace z_i \rbrace$ of the $\delta$-pulses and for $\lambda = 10$.
		%Panel (b):  The dependence of the asymptotic average velocity $\langle v \rangle$ on the mean value of shot-noise $\langle \eta(t) \rangle$ for three statistical distribution of the amplitudes $\lbrace z_i \rbrace$ of the $\delta$-pulses and for $\lambda = 4$.
		Panel (c): Asymptotic time-averaged transport velocity {\it vs.} the spiking rate $\lambda$ for three amplitude densities as depicted  in panel (a) and for an overall white shot noise intensity $D_P=10^{-3}$. The remaining  parameters are  as those given  in figure 3 and thermal fluctuation intensity $D_G = 0.001$.}
		\label{fig5}
\end{figure}

	We also address the dependence of  ANM transport for different statistics of the amplitude  $\lbrace z_i \rbrace$ entering generalized white shot noise.  In doing so we choose two additional amplitude statistics that derive from special cases of the Gamma distribution. Particularly, we study
	\begin{equation}
		\rho_2(z) = \zeta^{-2}\theta(z)z e^{-z/\zeta}
	\end{equation}
	and
	\begin{equation}
		\rho_3(z) = \frac{1}{2}\zeta^{-3}\theta(z)z^2 e^{-z/\zeta},
	\end{equation}
	where $\theta(z)$ is the Heaviside function. For the  density  $\rho_2(z)$, the first two moments read
	\begin{equation}
		\langle z_i \rangle = 2\zeta, \quad \langle z_i^2 \rangle = 6\zeta^2.
	\end{equation}
	As  a result,   upon inspecting equations  \eref{pwsnmom} and \eref{pwsnintensity}, the mean value is  $\langle \eta(t) \rangle = 2\sqrt{D_P\lambda / 3}$ and the white shot intensity reads $D_P = 3\lambda \zeta^2$. Likewise, for the  density  $\rho_3(z)$, we obtain
	\begin{equation}
		\langle z_i \rangle = 3\zeta, \quad \langle z_i^2 \rangle = 12\zeta^2.
	\end{equation}
In this case we find  $\langle \eta(t) \rangle = \sqrt{3 D_P\lambda/2}$ and $D_P = 6\lambda \zeta^2$. The main difference between the  exponential probability density $\rho_1(z)$ in equation (\ref{rho1}) and these  two integrable densities is a non-monotonic, bell-shaped form, see figure   5(a). As a consequence, with $\rho_1(z)$, very small noise amplitudes are the most  likely. In the case of $\rho_2(z)$,  the maximum  of the density  occurs for the amplitudes $z_i = \zeta$ while for    $\rho_3(z)$,  the amplitudes $z_i = 2 \zeta$  are  most probable. All three probability densities   are depicted in Fig. \ref{fig5}(a). Panel (b) of this figure depicts the dependence of the  averaged velocity $\langle v \rangle$ on the three statistical densities for  shot noise amplitudes $\lbrace z_i \rbrace$ at  a fixed spiking rate $\lambda = 10$.

We observe  that in the regime of ANM, white Poissonian noise with amplitudes density  $\rho_3(z)$ is slightly more effective. In Fig.  5(c) we show the behavior  for the various  amplitude statistics when the spiking rate
$\lambda$  is varied. Overall we find (for the chosen set of three examples) a weak dependence of the ANM regime on the statistics for the noise amplitudes. We therefore may assume that the statistics of the amplitudes only weakly impacts the overall ANM regime, apart from possibly some
cases with abnormal, stylized density features.

%%%%%%%%%%%%%%%%%%%%%%%%%%%%%%%%%%%%%%%%%%%%%%%%%%%%%%%%%%%%%%%%%%%%%%%%%%%	

\subsection{Comparison with deterministic bias}

\begin{figure}[h]
		\centering
		\includegraphics[width=0.49\linewidth]{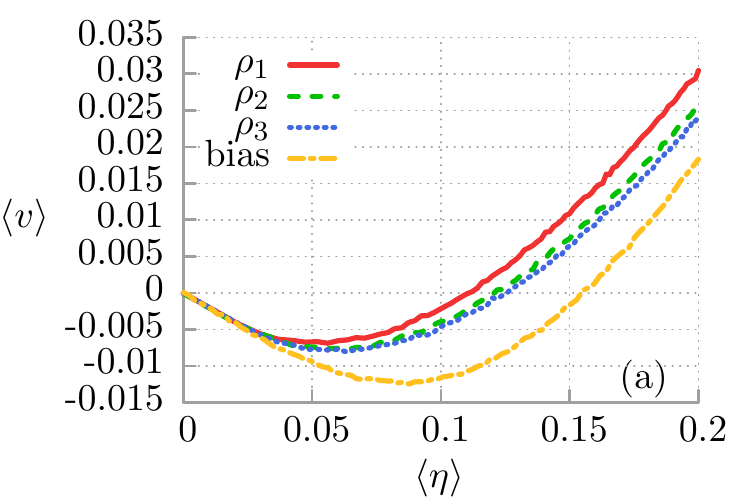}
		\includegraphics[width=0.49\linewidth]{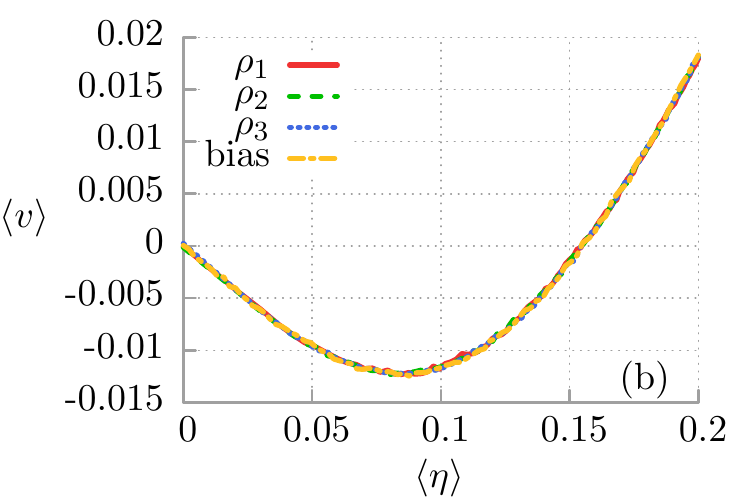}
		%\resizebox{0.49\linewidth}{!}{\input{tfig6a}}
		%\resizebox{0.49\linewidth}{!}{\input{tfig6b}}
		\caption{The dependence of the asymptotic time-averaged velocity $\langle v \rangle$ on the constant bias $f$ versus the asymptotic time-averaged  mean value of shot-noise $\langle \eta(t) \rangle$ for $\lambda = 4$ in panel (a) and  $\lambda = 512$ in panel (b). The thermal fluctuation intensity $D_G = 0.001$. The remaining  parameters are the same as in panel (b) of figure 2. Perfect equivalence (i.e. indistinguishable line plots)  of deterministic and random forcing is observed for a high spiking rate of $\delta$-pulses of Poissonian noise. }
		\label{fig6}
\end{figure}
	
As a final point of analysis, we compare ANM generated by Poissonian shot noise and the external force $F$ which is constant in space and time. So,  we have to consider the dimensionless form of the Langevin equation corresponding to 	 \eref{mini}, namely,
	\begin{equation}
		\ddot{x} + \gamma\dot{x} = - V'(x) + a\cos(\omega t) + f + \sqrt{2\gamma D_G}\, \xi(t),
	\end{equation}
	where the dimensionless deterministic constant force  $f = (L/\Delta V)\,F$.
%	Let us note that the case with a constant, deterministic bias can be obtained  for a spiking rate approaching $\lambda=\infty$ and fixed, non-random amplitude $z_i=F$.
 In order to compare the scenario of the deterministic force  $f$  with the system   driven by Poissonian white shot noise,  we need to impose the  additional condition
	\begin{equation}
		\langle \eta(t) \rangle = f.
	\end{equation}
We consider the following parameter regime:  the friction coefficient $\gamma = 0.9$,  the ac-driving amplitude   $a = 4.2$, the angular driving frequency   $\omega = 4.9$ and  the thermal fluctuation intensity $D_G = 0.001$. We stress that it is the same parameter regime as in figure 2(b), where the Ohmic-like transport is observed for the case of a low spiking rate $\lambda \le 0.3$. In figure 6, we show the dependence of the asymptotic average velocity $\langle v \rangle$ on the constant deterministic bias and mean value of shot-noise $\langle \eta(t) \rangle$  for the case of  (i) a moderate spiking rate
	$\lambda = 4$ and   (ii) a large spiking rate $\lambda = 512$.  For the low firing rate, one can observe normal transport; for  moderate values of the spiking rate $\lambda$ we detect small windows of occurrence of  ANM. Seemingly the case with   the deterministic bias is  most effective for ANM, yielding a wide regime of bias values $f$. For a large spiking rate of the white shot noise we indeed detect   the convergence towards the deterministic constant bias case, cf. the panel (b) in figure  \ref{fig6}.

%%%%%%%%%%%%%%%%%%%%%%%%%%%%%%%%%%%%%%%%%%%%%%%%%%%%%%%%%%%%%%%%%%%%%%%%%	

\section{Conclusions}
\label{sec4}
	With this work we presented a  detailed study of the transport properties of  an inertial Brownian particle which moves in a periodic, symmetric potential and which in addition is exposed  to periodic  harmonic ac-driving and (generalized) Poissonian white shot noise of finite bias $F$. We have demonstrated the possibility to manipulate the {\it direction} of transport just by adjusting the parameters of white shot-noise. Moreover, in such systems Poissonian white shot noise can induce  anomalous transport effects. In particular,  such dynamics is able to exhibit an absolute negative mobility regime. This ANM phenomenon has its roots in a pure-stochastic dynamics of the system and  is robust with respect to   the distribution of the random  amplitudes of the  $\delta$-pulses. In some regions of parameter space, one can find similar impact of Poissonian shot noise as by the deterministic bias. In general, the exact equivalence of both sources of bias 	
does not hold true.  However, in the ANM regime  the equivalence is observed   in the limiting case  of a spiking rate  $\lambda \to \infty$  of the  $\delta$-pulses.  For moderate-to-large $\lambda$,   the ANM induced by shot noise is suppressed as compared with the case of a deterministic bias. Notably, for a small spiking rate $\lambda$  the ANM response no longer is present and instead a normal, Ohmic-like behavior occurs.
 %What is even more surprising, properly prepared white shot-noise is equivalent  in this regime to constant deterministic force.
%Our results are  robust in the sense that the transport features are in essence  independent of the statistical properties of the random amplitudes for the $\delta$-spikes.

Our results  can readily be experimentally tested with an accessible setup consisting of a single resistively and capacitively shunted Josephson junction device operating in its classical regime.

\ack
The work supported in part by the grant N202 052940 (J.S. and J.L.)  and  the  ESF Program  "Exploring the Physics of Small Devices" (P.H. and J.L.).

\end{document}